\documentclass[12pt,nofootinbib]{revtex4}
\usepackage{amsmath}
\usepackage{amssymb}
\usepackage{epsfig}
\usepackage{tabularx}
\usepackage{graphicx}
\usepackage{inputenc}
\usepackage{url}
\usepackage{breqn}
\usepackage{hyperref}
\usepackage{footmisc}
\usepackage{verbatim}
\def\no{\tilde{\chi}^0_1}
\def\nt{\tilde{\chi}^0_2}
\def\nth{\tilde{\chi}^0_3}

\def\sb{\tilde{b}_1}

\begin{document}
\begin{titlepage}
\begin{center}
\hfill EFI-14-37
\end{center}
\vspace{.1cm}
\title{CMS kinematic edge from s-bottoms}
\vspace{1.5cm}
\author{\textbf{Peisi Huang$^{a,c}$ and Carlos E.M. Wagner$^{a,b,c}$} \\
\vspace{1.5cm}
\normalsize\emph{$^a$Enrico Fermi Institute \& $^b$Kavli Institute for Cosmological Physics,}\\
\normalsize\emph{University of Chicago, Chicago, IL 60637} \\
\normalsize\emph{$^c$HEP Division, Argonne National Laboratory, 9700 Cass Ave., Argonne, IL 60439}
\vspace{1.5cm}
}

\begin{abstract}
We present two scenarios in the Minimal Supersymmetric Extension of the Standard Model (MSSM) that can lead to an explanation of  the excess in the invariant mass distribution of two opposite charged, same flavor leptons, and the corresponding edge at an energy of about 78~GeV, recently reported  by the CMS collaboration. In both scenarios, s-bottoms are pair produced, and decay to  neutralinos and a b-jet. The heavier neutralinos further decay to a pair of leptons and the lightest neutralino through on-shell s-leptons or off-shell neutral gauge bosons.   These scenarios are consistent with the  current limits on the s-bottoms, neutralinos, and s-leptons. Assuming that the lightest neutralino is stable we discuss the predicted relic density as well as the implications for Dark Matter direct detection.  We show that consistency between the predicted and the measured value  of the muon anomalous magnetic moment may be obtained in both scenarios. Finally, we  define the signatures of these models that may be tested at the 13 TeV run of the LHC.
\end{abstract}
\maketitle
\end{titlepage}
\section{Introduction}
\label{section:intro}
After the Higgs discovery~\cite{AtlasHiggs,CMSHiggs}, the main goal of the LHC experiments is the search for new physics at the TeV scale.  Current searches at the 8 TeV LHC have provided no evidence of new physics beyond the Standard Model (SM).  There are, however, some intriguing signatures that may hint to the presence of new physics. For instance, in a recent analysis of the invariant mass distribution of two opposite charged, same flavor (SFOS) leptons~\cite{edge}, CMS has reported an intriguing excess of events with respect to the ones expected in the SM.  In this search  CMS looks for two isolated lepton final states using the 8 TeV data set with an integrated luminosity of 19.4 fb$^{-1}$. Events with  SFOS leptons are selected ($e^{+}e^{-}$ or $\mu^{+}\mu^{-}$) with both leptons having transverse momentum $p_T >$ 20 GeV and pseudorapidity $|\eta| <$ 2.4. CMS set additional requirements on jets and missing energy, and selects events with a number of jets $N_{jets} \ge$2 and missing transverse energy $E_T^{\rm miss} >$ 150 GeV or $N_{jets}\ge$3 and $E_T^{\rm miss}  > $100 GeV. The jets are required to have $p_T >$ 40 GeV and $|\eta| <$ 3.0.  The selected events are separated into a central signal region, where both leptons satisfy $|\eta| <$~1.4,  and a forward region, where at least one lepton satisfies 1.6 $< |\eta| <$ 2.4.  Then CMS performs a search for an edge in the invariant mass ($m_{ll}$) distributions by fitting the signal and background hypothesis to data in the range of 20 GeV $< m_{ll} <$ 300 GeV.  The best fit to the SFOS event distribution is obtained for an edge  at an energy of 78.7$\pm$ 1.4 GeV.  An alternative search is done by a counting experiment, without any assumption of the signal and background shape.  The counting experiment is performed in the mass range of 20 $< m_{ll} < $ 70 GeV, and an excess of 130$^{+48}_{-49}$ events are seen in the central region, corresponding to a local significance of 2.6~$\sigma$. 

In this article, we shall interpret the presence of this edge as a signature of the production of third generation supersymmetric particles at the LHC~\footnote{
During  the completion of this work,  an alternative  explanation of this kinematic edge in terms  of first and second generation s-quarks together with light s-leptons was presented~\cite{Allanach:2014gsa}.}. Supersymmetry is an attractive framework~\cite{Nilles:1983ge,Haber:1984rc,Martin:1997ns}, that leads to the unification of couplings at high scales and provides Dark Matter candidates in terms of the superpartners of the neutral Higgs and gauge bosons. Moreover,  for supersymmetric particle masses of the order of the TeV scale, low energy supersymmetry leads to the radiative breaking of the electroweak symmetry with a light, mostly SM-like Higgs boson with a mass which may be consistent with the value observed at the LHC~\cite{Mahmoudi:2012eh,Hall:2011aa,Heinemeyer:2011aa,Feng:2011aa,Carena:2011aa,Draper:2011aa}. It has been pointed out that the study of kinematic  edges can be important for the detection of light supersymmetric particles (for recent work, see~\cite{Dreiner:2010gv,Lester:2005je}), and several kinematic variables have been proposed to distinguish  new physics from background. In this article, we shall focus on the edge in the $m_{ll}$ distribution. 

The kinematic edge can be explained by the presence of a light s-bottom~\footnote{The explanation of the edge events with via the decay of s-bottoms is consistent with the benchmarks proposed by CMS in Ref~\cite{edge}. However, the CMS benchmarks, based on the heavier neutralinos decay to an off-shell Z and the lightest neutralino, are in tension with the ATLAS limits on b-jet + $E_T^{miss}$ channel.}. The s-bottoms are pair produced in the detector, and one of the s-bottoms decays to a b-jet and the LSP. The other one decays to a heavier neutralino and a b-jet, and the heavier neutralino subsequently decays to the lightest neutralino and a pair of SFOS leptons through a s-lepton or an off-shell Z, depending on the mass of the s-leptons and the 
composition of the neutralinos.  We therefore defined two scenarios. In scenario A, the s-bottom decays  through the decay chain of $\sb \rightarrow b \nt \rightarrow b \tilde{l}l \rightarrow b l^+l^-\no$.  In scenario B, instead, the s-bottom decay features the cascade decay of $\sb \rightarrow b \tilde{\chi}_{2,3}^0 \rightarrow b Z^* \no \rightarrow b l^+ l^- \no$.  
Thus, the two s-bottoms together give 2 b-jets, a pair of SFOS leptons and  missing energy. Under these conditions, the $m_{ll}$ distribution will feature a kinematic edge. 

This article is organized as follows : In section~\ref{section:s-bottom}, we  will analyze the two possible scenarios in more detail, and study the resultant contributions to the kinematic edge. 
 We also discuss the implications of these scenarios for Dark Matter and the muon anomalous magnetic moment. In section~\ref{section:constraints} we consider possible constraints on both scenarios from the LHC. We reserve section~\ref{conclusions} for our conclusions.

\section{s-bottom contributions to the dilepton kinematic edge}
\label{section:s-bottom}

Scenarios with pair produced s-bottoms are able to explain the kinematic edge. The spectra in Fig~\ref{fig:slep} and Fig~\ref{fig:Z} are examples of possible scenarios, where  the particle masses necessary to explain the data in each of these scenarios are presented in Table~\ref{table:spectra}. 
\subsection{Scenario A}
\begin{figure}[tbh]{
\includegraphics[width = 8cm,clip]{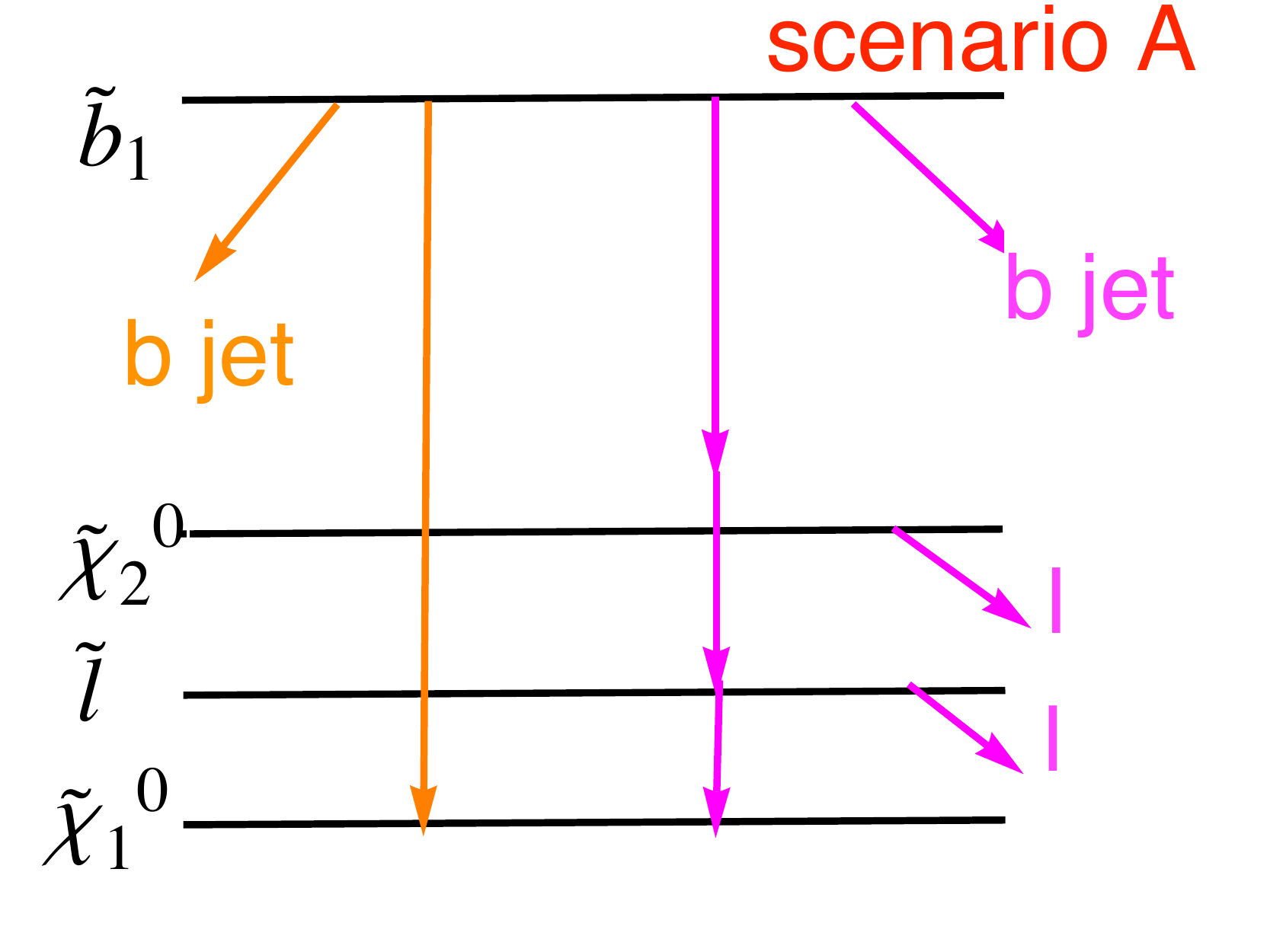}
\caption{A spectrum that could account for the dilepton kinematic edge. $\nt$ decays to the LSP and a pair of same flavor opposite sign dileptons through a light s-lepton.}
\label{fig:slep}
}
\end{figure}

The spectrum in Fig~\ref{fig:slep} features a s-bottom with a mass around 390~GeV and a light s-lepton. The s-bottom can decay to a $\nt$ and a b-jet. The $\nt$, with a mass around 340 GeV can decay to two leptons and a $\no$ through a right-handed s-electron or a s-muon with  masses $m_{\tilde{e}_R} = m_{\tilde{\mu}_R} = m_{\tilde{l}}$ around 300~GeV.  The mass of the LSP is chosen to be 260 GeV.  Those two leptons will have same flavor and opposite signs, and the edge of the invariant mass of the dilepton will be at
\begin{equation}
m_{ll}^{edge} = \sqrt{\frac{(m_{\nt}^2-m_{\tilde{l}}^2)(m_{\tilde{l}}^2-m_{\no}^2)}{m_{\tilde{l}}^2}}, 
\label{edgesleptons}
\end{equation}
which is about 80 GeV in this spectrum. In Eq.~(\ref{edgesleptons}) $m_{\tilde{l}}$ is the s-lepton mass and $m_\no$ and $m_\nt$ are the lightest and second lightest neutralino masses, respectively. The competing decay channel of the s-bottom is a b-jet and the LSP.  Therefore, the pair produced s-bottoms, with one s-bottom decaying to a b-jet and the LSP, and the other decaying through the decay chain discussed above, will contribute to the SFOS dilepton + $\ge$2 jets + missing energy channel with a kinematic edge around 80 GeV. Also, since the s-bottom decays to either a b-jet and missing energy, or a b-jet, two leptons and missing energy, there will be no significant contributions to the 2b-jets plus 2 jets channel from s-bottom pair production. 

\begin{table}
\centering
\begin{tabular}{c|c|c}
\hline
parameter & scenario A & scenario B  \\
\hline
$m_{\sb} $ (GeV) & 390  &  330 \\
$m_{\no}$ (GeV)& 260  & 212 \\
$m_{\nt}$ (GeV) & 340 &288 \\
$m_{\nth}$ (GeV)& $\sim$ 500 & 290 \\
$m_{\tilde{l}}$ (GeV)&  297 &500 \\
$\tan\beta$ & 25 & 50 \\
\hline
$\sigma(p p \rightarrow \sb \sb) $ (pb)  & 0.42 & 1.14 \\
BF($\sb \rightarrow b \no$) & 0.93 & 0.56 \\
BF($\sb \rightarrow b\nt$) & 0.07  &0.25 \\
BF($\sb \rightarrow b\nth$) & 0  &019  \\
$\Delta a_{\mu}$ & 2.0$\times 10^{-9}$ &  2.7 $\times 10^{-9}$\\
$\Omega h^2$  &0.11 & 0.11\\
$\sigma_{SI}^{p}$ in cm$^{2}$ &4$\times$ 10$^{-45}$  & 2.7$\times$10$^{-44}$ \\
\hline
\end{tabular}
\caption{Parameters for the two scenarios described in the text.}
\label{table:spectra}
\end{table}
The mass parameters in scenario A were chosen in order to give a sufficiently large cross-section without being in conflict with other experimental constraints, which will be discussed in more details in section~\ref{section:constraints}. We choose $m_{\sb}$ around 390 GeV so that it has a sizable production cross section.  The mass of the lightest neutralino is chosen to be sufficiently low to allow the existence of an edge, but not low enough to lead to a conflict with searches for pair production of $\sb$ in the 2~b-jets plus $E_T^{\rm miss}$ channel.  The mass of the second lightest neutralino is chosen to lead to the required edge while avoiding a degenerate spectrum which will give too soft b-jets.
For a $\sb$ of 390 GeV a decay branching ratio $BF(\sb \to b \no) = 1$, the bound on the LSP mass is 260 GeV, and in order satisfy the above requirements we chose the $\no$ mass to be close to this value, which is consistent with the ATLAS experimental bounds due to the fact that in scenario A $BF(\sb \to \no) < 1$.  In order to induce a large excess of events in the  2~b-jets plus SFOS leptons channel without generating a similarly large number of events in the 2-b-jets plus four leptons channel, we require that $BF(\sb \to b  \nt) \ll BF(\sb \to b \no)$.  A simple way of satisfying this requirement is to assume that $\sb$ is mostly right-handed, $\nt$ is mostly a Wino and $\no$ is mostly a Bino. $\tilde{\chi}_1^{\pm}$ is wino like. Observe that the chargino contribution to the sbottom decay branching fraction,  BF($\sb \rightarrow t^{*} \tilde{\chi}_1^{\pm}$), is highly suppressed due to phase space factors. 

The parameters are further constrained by the requirement of obtaining a proper Dark Matter relic density and a value of the anomalous magnetic moment of the muon consistent with experiment. 
 To obtain the right relic density, we can either mix the bino-like LSP with Higgsinos, so that the LSP can annihilate more efficiently through the Higgsino components, or approach the so-called A-funnel region, where $m_{\no} \simeq m_A/2$, with $m_A$ being the CP-odd Higgs mass, so the LSP can annihilate efficiently through the resonant mediation of the heavy Higgs bosons. For a neutralino mass $m_\no \simeq 260$~GeV, the A-funnel region is excluded by the CMS Higgs to $\tau\tau$ searches~\cite{htautau} for $\tan\beta > 20$. Then, for large $\tan\beta$ the right relic density requires a small Higgsino mass parameter $\mu$ to get a large enough Higgsino component. 
 A small $\mu$ and a large $\tan\beta$ make the Higgsino components in $\nt$ too large to give a BF($\sb \rightarrow \nt b$) small enough to be consistent with the 4 lepton + missing energy searches. That means that the right relic density and a small BF($\sb \rightarrow \no b$) together favor  moderate values of $\tan\beta$.  

At the same time, to get the right muon anomalous magnetic moment, a larger $\tan\beta$ is favored. So if we put the restriction on the relic density, BF($\sb \rightarrow \no b$) and muon g-2 together, a sizable, but not very large value of $\tan\beta$ is favored. When $\tan\beta = 25$, as in scenario A, the bound on $m_A$ from the CMS Higgs to $\tau\tau$ searches is about 600 GeV. Then, values of $\mu$ around 500 GeV are necessary to give the right relic density.  Also, BF($\sb\rightarrow \no b$) is 7.4$\%$ for $\tan\beta = $ 25 and $\mu =$ 500 GeV, which is consistent with the constraints coming from 4 lepton searches, which we are going to discuss  in detail in section~\ref{section:constraints}.  Then we choose the mass of the right-handed sleptons to be 320 GeV to give the $m_{ll}$ edge at 80 GeV according to Eq~(\ref{edgesleptons}). In addition, we chose the mass of the left-handed s-leptons to be around 400 GeV to give a larger contribution to the muon anomalous magnetic moment, which in this scenario is of about $\Delta a_{\mu} =$ 2 $\times 10^{-9}$. This is consistent with the experiment value $\Delta a_{\mu} = (2.87 \pm 0.8) \times$ 10$^{-9}$~\cite{Barbieri:1982aj,Ellis:1982by,Kosower:1983yw,Moroi:1995yh,Carena:1996qa,Czarnecki:2001pv,Feng:2001tr,Martin:2001st,Freitas:2014pua}. 
Smaller values of $\mu$ would further enhance the value of  $\Delta a_{\mu}$, but they will also lead to an enhancement of the BF($\sb \rightarrow b \nt$) and therefore we will not consider them in this analysis.

\begin{figure}[tbh]{
\includegraphics[width = 10cm,clip]{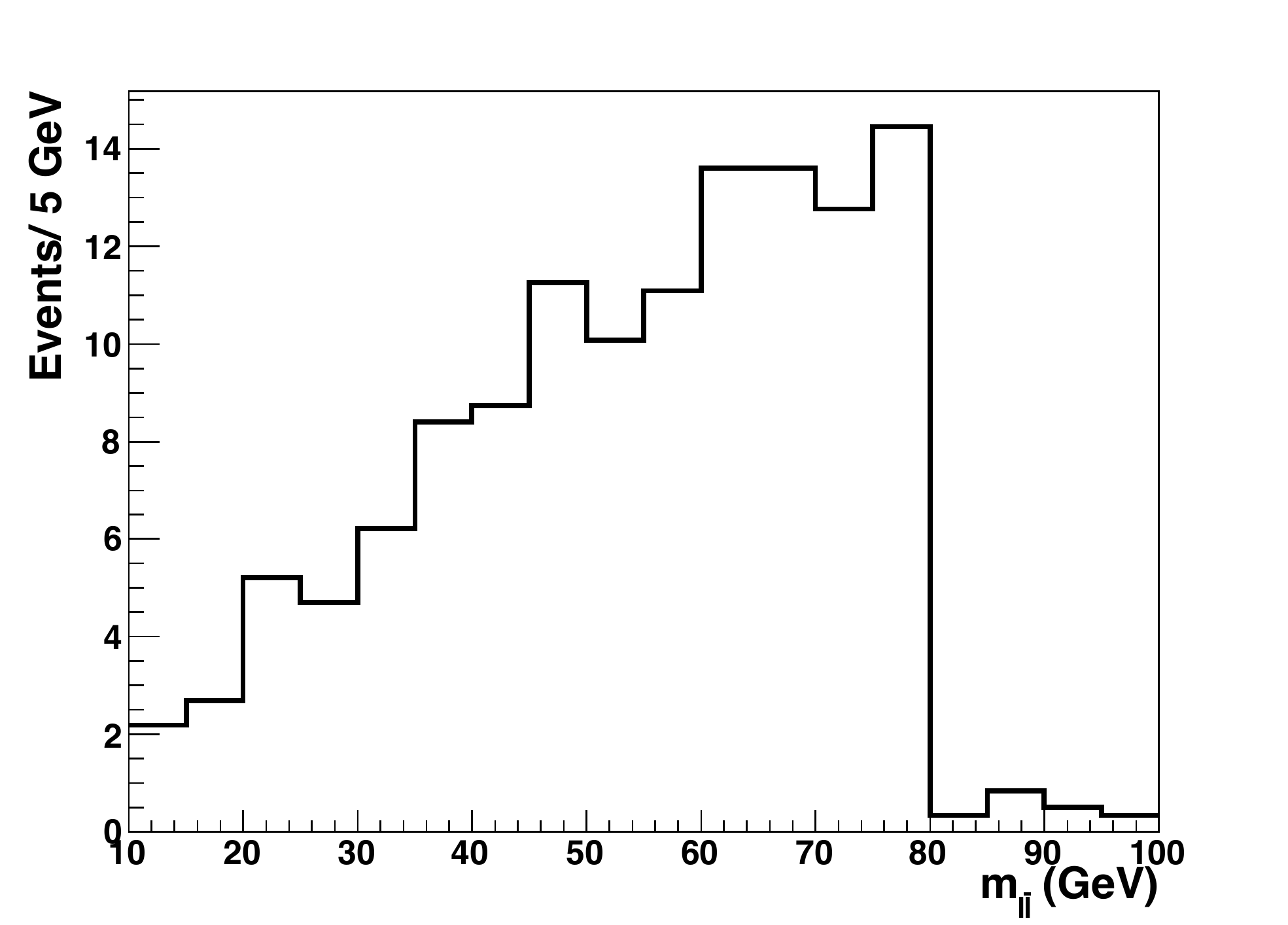}
\caption{Invariant mass distribution of the same flavor, opposite sign dileptons. See the spectrum in Fig~\ref{fig:slep} and Table~\ref{table:spectra}.}
\label{fig:edge}
}
\end{figure}

We performed a collider study of s-bottom pair production using the spectrums shown in Fig~\ref{fig:slep}.  We generated the events with Isajet~\cite{isajet}, and pass the event to Pythia~\cite{pythia} and PGS~\cite{PGS} for showering and the detector simulation and follow the CMS event selection described in section~\ref{section:intro}. The production cross section was scaled to the Prospino NLO results~\cite{prospino}. In the scenario A under analysis, $\nt$ is Wino-like and $\no$ is Bino like, BF($\sb \rightarrow \nt b$) is $7.4\%$, BF($\sb \rightarrow \no b$) is 92.6$\%$ and $\sigma( p p \rightarrow \sb \overline{\sb}) = 0.42$ pb.   We found out that at the 8 TeV LHC, with a total integrated luminosity of 19.4 fb$^{-1}$,  110 SFOS dilepton events are expected in the  central signal region and 13.4 events in the forward signal region. As stressed before, there is no significant contributions to the $\ge$2bjets + $\ge$2 jets channel. The invariant mass distribution of the dilepton system shows an edge at about 80 GeV, as predicted by Eq.~(\ref{edgesleptons}), and  shown in Fig~\ref{fig:edge}.

 Although we have not attempted to find the optimal values of the supersymmetric particle masses consistent with the observed signal, we have analyzed the effects of possible variations of these parameters. 
 A s-bottom lighter than 390 GeV will have a larger production rate, but the generation of the kinematic edge becomes more difficult due to the fact that for a BF($\sb \to \no b) \simeq 1$ the experimental bounds
on the LSP mass for s-bottoms with masses between 300 GeV and 400 GeV (which will be discussed in more detail in the next section)  do not change much with the mass of s-bottoms. That means, for a lighter s-bottom, the allowed mass for the LSP would only be slightly smaller than 260 GeV.   Therefore,  in order to reproduce the edge the heavier neutralino mass should be kept at approximately the same value and the BF($\sb \rightarrow b \nt$) would be smaller than in the example we discussed above. For instance, for a s-bottom around 385 GeV, the cross section is increased to 0.45 pb while BF($\sb \rightarrow b\nt$) is reduced to 0.064. Assuming the kinematic properties do not change in a significant way, the number of events expected would be rescaled by the ratios of the corresponding cross sections and branching ratios, implying about 103 events in the central region.   Hence, a  small reduction of the signal would be obtained.  

If the mass of s-bottom is larger than what we discussed in the scenario A example, the s-bottoms pairs will be produced at a lower rate. 
For a 400 GeV s-bottom, for instance, the production cross section drops to 0.36~pb. The number of SFOS events 
may be  recovered by increasing BF($\sb \rightarrow b\nt$) (for instance increasing slightly the Higgsino composition of $\nt$ or changing the masses of $\nt$ and of the LSP), a possibility which  is however constrained by the searches in the 4 lepton channel. A more detailed collider study would be needed to estimate the largest BF($\sb \rightarrow b\nt$) allowed by the 4 lepton searches, which is beyond the discussion of this article, but $\sb$ masses above 400~GeV are significantly constrained by this requirement. 
 
As we explained above, the value of $\tan\beta$ was fixed from the requirement of consistency with Dark Matter and muon anomalous magnetic moment constraints.  The predicted Dark Matter relic density in Scenario A, for $m_A \simeq 600$~GeV, is $\Omega h^2 \simeq$ 0.11, which is consistent with the cosmological observations $\Omega h^2 = $0.1198 $\pm$0.0026\cite{Plank}. The value of $\tan\beta$ may be lower, implying a lower
bound on the CP-odd Higgs mass $m_A$.   Therefore, we can go to the A-funnel region (resonant annihilation through the CP-odd Higgs) to get the right relic density and $\mu$ can be chosen to the value that gives a BF($\sb\rightarrow \no+b$) large enough to contribute about 130 event in the edge and small enough to be consistent with the 4 lepton searches. 
However,   since the main contribution to the muon $g-2$ in the MSSM is amplified by $\tan\beta$, the value of $\Delta a_{\mu}$ would be smaller than the one obtained in scenario A. 
In the region where $\tan\beta$ is large, instead, the Higgs to $\tau\tau$ searches prevents $m_A$  to approach the A-Funnel region. 
At the same time, a large $\tan\beta$ means $\nt$ must be very wino-like so that the BF($\sb \rightarrow b\nt$) can be small enough to not let the s-bottom signals show up in the 4 lepton channel. That means $\mu$ is also required to be large in this region. Then the LSP is Bino-like, and away from the A-funnel region, so the relic density is larger than the observed value. 
For instance, for $\tan\beta$ = 50, and $m_A$ = 1 TeV, the predicted Dark Matter relic density in Scenario A is $\Omega h^2 \simeq$ 0.8, which is higher than the cosmological observations~\footnote{The over-abundance problem can be solved, for instance, by introducing a scalar field $\phi$, which dominates the energy density of the early universe before the nucleosynthesis era~\cite{Gelmini:2006pq}.  
For example, for a number of neutralinos produced per $\phi$ decay and per 100 TeV mass of $\phi$, $\eta \le$ 10$^{-6}$,  and a reheating temperate of  about 1~GeV an observable relic density consistent with experiment may be obtained.}. 
In general, for values of $\tan\beta$ significantly larger or smaller than the one chosen in scenario A there will be tension between the requirement of obtaining a value $\Delta a_{\mu}$ consistent with the observed experimental value and the obtention of the proper relic density. 
 
The spin-independent cross section of the LSP scattering off a proton for Scenario A is $\sigma_{SI}^p = $3.6$\times10^{-45} {\rm cm}^2$, which is consistent with LUX~\cite{LUX} and will be probed by Xenon1T and other future experiments~\cite{Xenon1T}. If the sign of $\mu \times M_1$ is flipped, but the rest of the parameters are kept as  defined in scenario A (with only small variations to obtain the proper relic density) then $\sigma_{SI}^p$ is reduced to about $3 \times10^{-47}$ cm$^2$ due to the destructive interference between the contribution from the SM-like Higgs and the heavy CP-even Higgs~\cite{Ellis:2000ds,Baer:2006te,PhysRevD.63.065016,BS}.  Xenon1T might be able to probe this scenario.  In order to keep a positive contribution to $\Delta a_{\mu}$, however, $\mu \times M_2$ should be kept positive.  For negative values of $\mu \times M_1$ and positive 
values of $\mu \times M_2$, the contribution to $\Delta a_{\mu}$ is reduced by about 20\% because the sign of  right-handed s-lepton and neutralino loop contribution turns negative, but  the dominant contributions from the left-handed s-lepton and neutralino  as well as the  s-neutrino and chargino loop contributions remain approximately the same.

\subsection{Scenario B} 

\begin{figure}[tbh]{
\includegraphics[width = 8cm,clip]{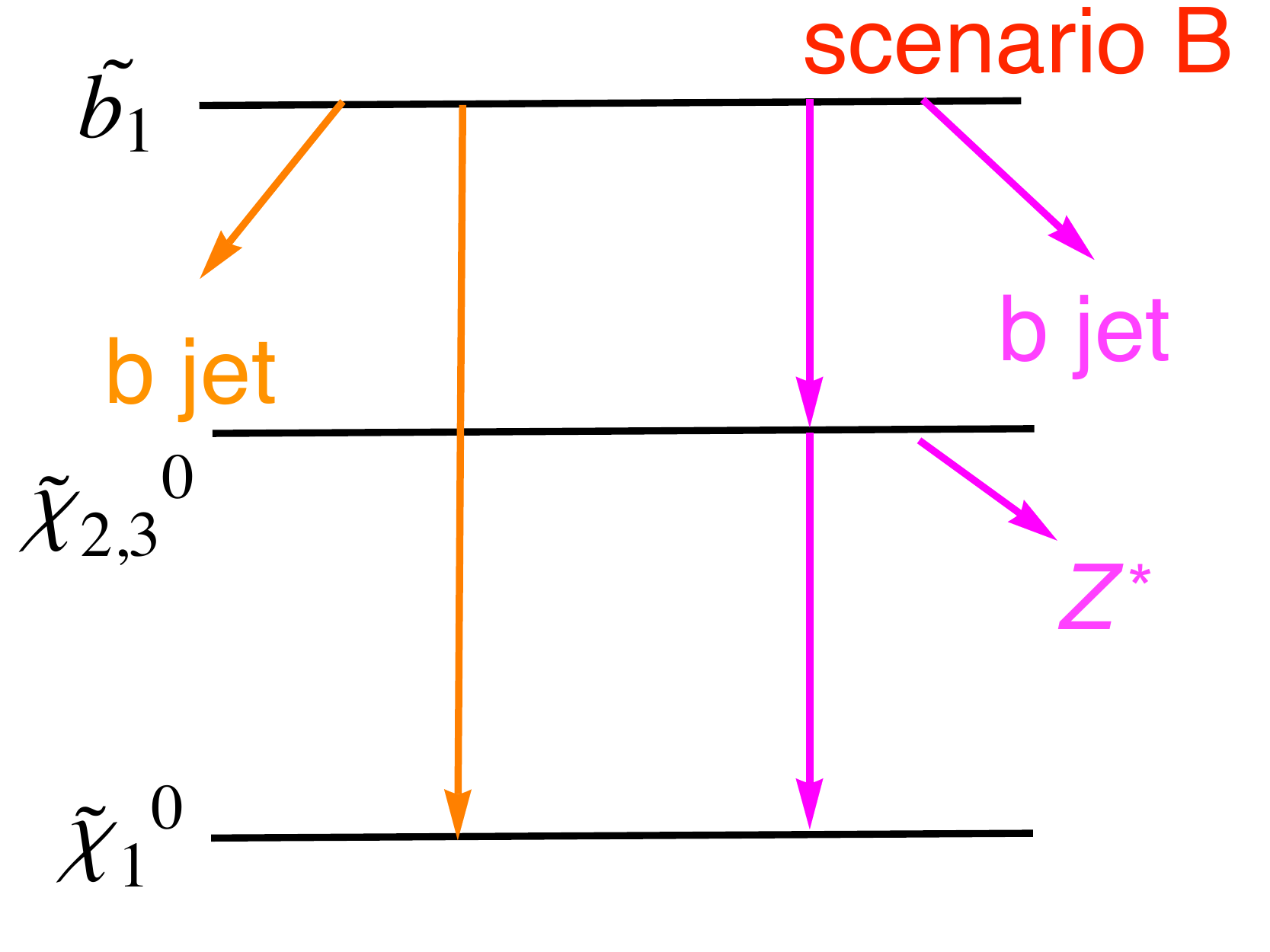}
\caption{A spectrum that could account for the dilepton kinematic edge. $\nt$ and $\nth$ decay to the LSP and a pair of same flavor opposite sign leptons through an off-shell Z.}
\label{fig:Z}
}
\end{figure}

An alternative scenario will be that the heavier neutralino decays to a $\no$ and an off-shell Z, then the Z$^{*}$ decays to a $l^{+}l^{-}$ pair. A spectrum that could be responsible for the CMS kinematic edge is shown in Fig ~\ref{fig:Z}. Contrary to what happens in scenario A, since the branching fraction of a Z$^{*} \rightarrow l^{+} l^-$ is about 6 $\%$, (summed over electrons and muons), a sizable $BF(\sb \rightarrow \nt  b)$ is needed to get the number of events around 100 without lowering the $\sb$ mass to the current excluded region. 
To fulfill  that requirement, we have $\nt$ and $\nth$ Higgsino-like, and $\tan\beta$ large to enhance the $\sb\widetilde{H}b$ coupling.  Since the BF($\sb \to b \no$) is significantly lower than one, the mass of the $\sb$ could
be lower without leading to a conflict with the 2~b-jets  + $E_T^{\rm miss}$ data.

In the spectrum shown in Fig~\ref{fig:Z}, we choose the $\sb$ mass around 330~GeV, the Higgsino mass parameter $\mu$ to be around 290~GeV and the LSP mass at 212~GeV. Then for these values of the mass parameters, $BF(\sb \rightarrow \nt  b) = $~0.25 and $BF(\sb \rightarrow \nth b) =$~0.19.  In this case, $BF(\sb \rightarrow \no l^+ l^- b)$ is around 0.44 $\times 0.06 \simeq$~0.03, which is small enough to suppress the 4 lepton mode and large enough to contribute about 100 events to the dilepton edge. In this scenario, the s-bottom pair production will also contribute to the $\ge$2 jets + $\ge$2 b-jet channel.  There is a potentially large signal in this channel coming from s-bottom pair production, and in the next section we shall discuss the constraints coming from it. 
Let us only emphasize here that the jets coming from the heavier neutralino decays tend to be soft and there are over-whelming backgrounds associated, for instance, with $t\bar{t}$ production, so that the scenario B is still consistent with this constraint.

In scenario B, the production cross section is $\sigma\left( p p \rightarrow \sb \overline{\sb}  \right) = $ 1.14~pb. At the 8 TeV LHC,  with a luminosity of 19.4~fb$^{-1}$, there are 80 events in the central signal region and 9.0 events in the forward signal region. In this case, the predicted edges are located at the mass difference between the heavier and the lightest neutralino,
\begin{equation}
m_{ll}^{edge} = m_{\nt,\nth} - m_{\no}, 
\label{edgenos-leptons}
\end{equation}
that was chosen to be  78~GeV and 76~GeV for the third and second lightest neutralino in this scenario, respectively.  

\begin{figure}[tbh]{
\includegraphics[width = 8cm,clip]{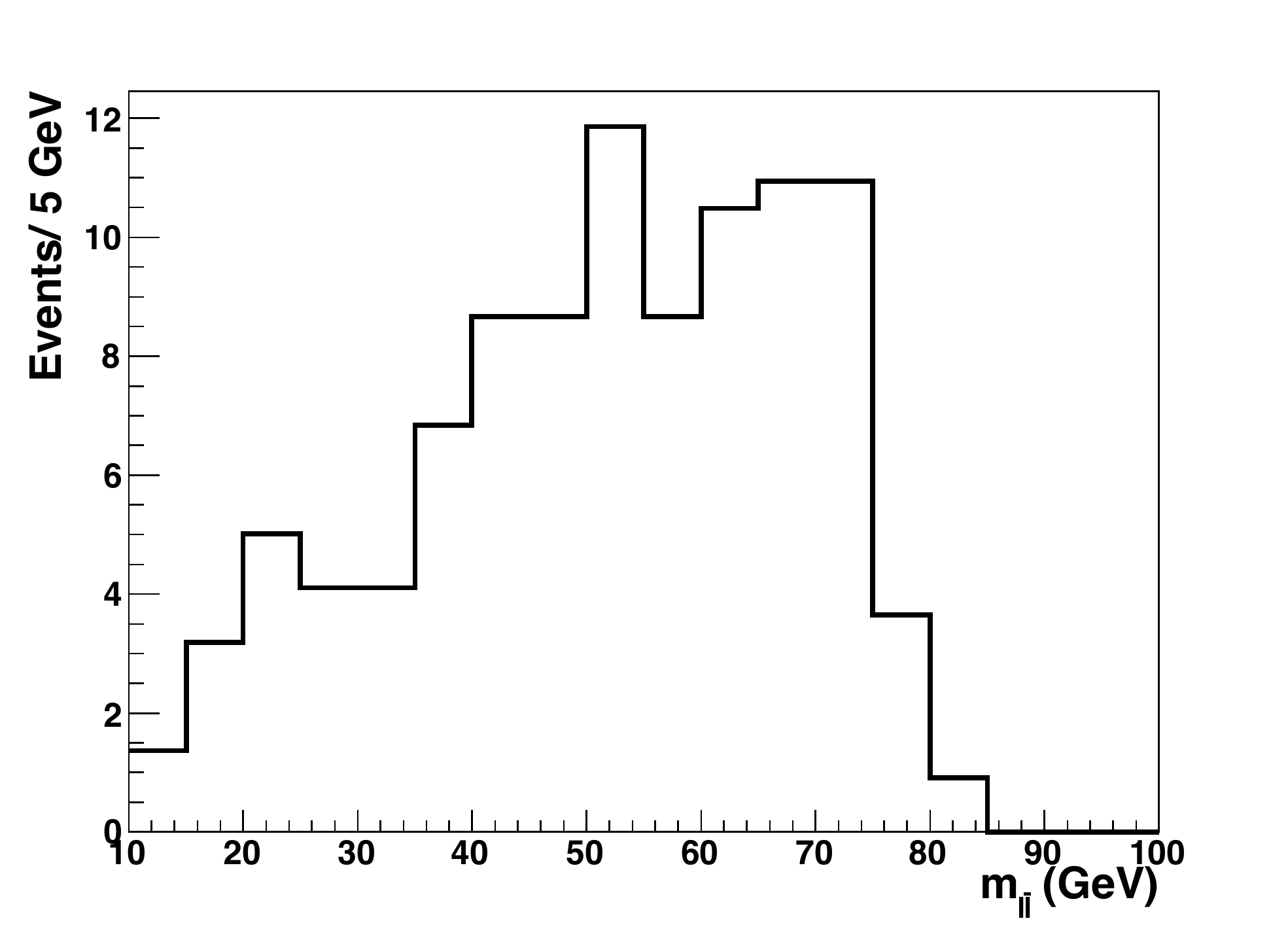}
\caption{Invariant mass distribution of the same flavor, opposite sign dileptons.  The edge is located at  the mass difference between the Higgsinos and $\no$. See the spectra in Fig.~\ref{fig:Z} and Table~\ref{table:spectra}. }
\label{fig:edgeZ}
}
\end{figure}

These edges can be seen in the invariant mass distribution presented in  Fig~\ref{fig:edgeZ}.  At the current luminosity, the mass splitting between the two Higgsinos $m_{\nth} - m_{\nt} \sim$ 2 GeV is sufficiently small to not be distinguishable in this distribution, and a single edge appears at about 78~GeV, consistent with the CMS data.  

The number of events obtained in the central region in this scenario is lower by about one standard deviation than the central value reported by CMS.  This amount may be increased by selecting a larger value of  BF($\sb \to b \nt,\nth$) or  a larger $\sb$ production cross section, associated with a lower $\sb$ mass.  However, any increase on these quantities will also lead to an increase of the number of events in the $\ge$2 jets, $\ge$2 b-jets + $E_T^{\rm miss}$ channel (as well as more events in the four lepton + $E_T^{\rm miss}$ channel). As we shall discuss in section~\ref{section:constraints},  there is already a slight tension between the predicted number of events in scenario B and the 
ones observed by the LHC experiments.  Hence,  it is in general difficult to obtain a significantly larger number of SFOS dilepton in this scenario.  

Moreover, 
%
for a lighter s-bottom,  the production cross section increases but BF($\sb \rightarrow b \tilde{\chi}^0_{2,3}$) decreases. For instance, for $m_\sb = $320 GeV, BF($\sb\rightarrow b\tilde{\chi}^0_{2,3}$) $\sim$ 0.35 and the cross section is 1.37~pb, which rescaling the previous result by the ratio of the corresponding production cross sections and branching ratios leads to about 88~events.  However, the bottom quarks coming from the $\sb$ decay 
are softer implying that the actual number of events in the edge remains approximately the same as for $m_\sb = 330$~GeV.  This could be improved by lowering the overall neutralino mass scale, but then the constraints from s-bottom direct detection in the 2~b-jets + $E_T^{\rm miss}$ channel become stronger, making an increase of events difficult (beyond the problems in the $\ge$2jets, $\ge$ 2~b-jets + $E_T^{\rm miss}$ channel mentioned above).   
Similar considerations apply for a heavier s-bottom.  Although in such a case the production cross section decreases, a larger BF($\sb\rightarrow b\tilde{\chi}^0_{2,3}$) is expected, together with harder b-jets and missing energy.  Therefore, there may be some space with a heavier s-bottom to obtain roughly 100 events in the central region, despite the production cross section being smaller.  Again, one of the main constraints in the region with a heavier s-bottom is from the $\ge$ 2-jet $\ge$ 2 b-jet channel,  with number of events in this channel scaled up by a factor of BF($\sb\rightarrow b\tilde{\chi}^0_{2,3}$)$^2$. 
%
We will come back to these issues in the next section.

This scenario leads to a proper anomalous magnetic moment of the muon for natural values of the s-lepton and Wino masses.  In order to  accommodate the muon $g-2$ result, the s-leptons should not be too heavy. For instance, for $m_{\tilde{l}_R} = m_{\tilde{l}_L}$ =~500 GeV, $\rm{M}_2 \simeq$ 600 GeV and heavy left-handed sleptons $\Delta a_{\mu} =$ 2.7 $\times 10^{-9}$, which is  consistent with the experiment value  $\Delta a_{\mu} = (2.87 \pm 0.8) \times$ 10$^{-9}$. 

Contrary to what happens in scenario A, the presence of light Higgsinos and Binos in the spectrum makes it possible to obtain a large enough Dark Matter annihilation cross section to ensure consistency with the observed relic density without the need
of any fine tuning of parameters.  
The predicted Dark Matter relic density for Scenario B is $\Omega h^2 \simeq$ 0.11, assuming the CP-odd Higgs is heavy, $m_A \simeq 1.5$~TeV, which is consistent with experiments. The spin-independent cross section of the LSP scattering off a proton is $\sigma_{SI}^p = $2.7$\times10^{-44} {\rm cm}^2$, which is consistent with LUX~\cite{LUX} and will be probed by Xenon1T and other future experiments~\cite{Xenon1T}. As discussed above, if the sign of $\mu \times M_1$ is flipped, then due to the destructive interference between the light and heavy CP-even Higgs bosons, $\sigma_{SI}^p$ is reduced to $\sim 1 \times10^{-46}$~cm$^2$ and Xenon1T may be able to probe this scenario.  Provided the sign of $\mu \times M_2$ remains positive, the contribution to $\Delta a_{\mu}$ is reduced by about 10\% in this case.

 \section{constraints from LHC searches and possible future searches}
\label{section:constraints}

As we discussed in the previous section, there are several constraints on the presence of light s-bottoms and  neutralinos coming from both ATLAS and CMS. The constraints from direct searches for s-bottoms decaying into bottom quarks and missing energy~\cite{CMS_sb,ATLAS_sb}~apply here. For scenario A, the direct limit for s-bottoms shows that, for a mass of the LSP of about 260 GeV, and a BF($\sb \rightarrow b \no) = 1$, s-bottoms with masses from 390 GeV to 620 GeV are excluded by ATLAS, while CMS has a weaker limit. Due to the smaller BF($\sb \to b \no)$, these bounds are weakened for this scenario.  In particular, as we stressed in the previous section, since we chose masses close to the ATLAS limit for a BF($\sb \to b \no) = 1$,  the mass parameters in scenario A are beyond the  current ATLAS limit.

For scenario B, the direct limit for s-bottoms in the 2~b-jet + $E_T^{\rm miss}$ channel show that for an LSP mass about 210 GeV, s-bottoms with masses from 240 GeV to 655 GeV are excluded if BF($\sb \rightarrow \no b$) =1. 
In scenario B, BF($\sb\rightarrow \no b$) is about 0.56, so the limits are weakened. We studied the number of events expected in this scenario to compare with the ATLAS result~\cite{ATLAS_sb}. In the ATLAS analysis, events are separated into two signal regions. In SRA, a large mass splitting between $\sb$ and $\no$ is expected, identifying two b-tagged high $p_T$ jets as products of the two s-bottom decays. Any other central jets or leptons are vetoed. SRB, instead, targets signal events with small mass splitting, by selecting events with a high $p_T$ jet, which likely be produced as initial state radiation, recoiling against the s-bottom pairs. The two additional jets are required to be b-tagged and large missing energy E$_T^{\rm{miss}} >250$ is imposed.  We expect 23.4 events in SRA, which is consistent with ATLAS results -- ATLAS expected 157.2$\pm$ 14.6 and observed 174 events. The number of events in separate bins is also consistent with ATLAS analysis. We expect $<$ 0.46 events, while ATLAS observed 1 events in SRB. With more integrated luminosity, we expect to see an excess in the 2~b-jet + $E_T^{\rm{miss}}$ channel.  

Both scenarios contribute to the 4-lepton, $\ge$ 1 jets plus missing energy channel. CMS studied this channel~\cite{CMS_4lep}.  CMS does not observe any events in the region $H_T >$200 GeV and $E_T^{\rm miss} <$ 50 GeV, while 0.08 events were expected in the SM. In the region of $H_T < $ 200 GeV and $E_T^{\rm miss} <$  50 GeV, 1 event is observed while 0.23 events are expected in the SM.   

We studied the expected number of four lepton events in our scenarios. In Scenario A,  0.3 events are expected in the $H_T >$200 GeV and $E_T^{\rm miss} <$ 50 GeV region and 1.6 events are expected in the $H_T <$200 GeV and $E_T^{\rm miss} <$ 50 GeV region. For scenario B, the expected signals are 0.9 events and 1.4 events in the $H_T >$ 200 GeV and $H_T <$ 200 GeV region, respectively. Both scenarios agree with the CMS results, and with more integrated luminosity, we expect  to see an excess in the 4-lepton, $\ge$ 1jets plus missing energy channel. 

A strong constraints to Scenario B is associated with the searches in $\ge$ 2 jets + $\ge$ 2 b-jets channel.  About 200 events with $E_T^{\rm miss} >$ 150 GeV are expected in this scenario.  We compare the predicted
number of events with the observations of the LHC experiments in the $\ge$~4-jets + $E_T^{\rm miss}$ channel~\cite{ATLAS_gl}. For loose cuts the number of observed events is consistent with the ones associated with s-bottom production. For tight cuts, the number of observed events is small but it falls short of the ones expected in the SM. Therefore, there is a small tension with the numbers predicted from s-bottom production in scenario B, which however are still much smaller than background.  Therefore, if the edge is confirmed, we expect this channel to provide a relevant test of scenario B at the next run of the LHC.


 Also, the limits from chargino and neutralino searches should be taken into consideration~\cite{CMS_EW,ATLAS_EW}. The limits are quite weak for a LSP as heavy as 200 GeV, and our choice of parameters in section~\ref{section:s-bottom} is well allowed by both experiments. Both scenarios contribute to the single lepton channel, but both experiments require more than 3 hard jets in the single lepton search~\cite{CMS_singlelep,CMS_singlelep2,ATLAS_singlelep,ATLAS_singlelep2} and therefore do not set any constraints in our scenarios from  this channel. Scenario A also gets constrained by direct s-lepton searches, but the current limits from direct s-lepton searches do not cover the LSP mass larger than 100 GeV~\cite{ATLAS_EW,CMS_EW} assuming the lighter s-lepton is right-handed.    
%

 \section{conclusions}
 \label{conclusions}
To summarize, we have shown two possible scenarios with s-bottoms with masses around 300 to 400 GeV to explain the recent CMS excess of events in the SFOS dilepton and missing energy channel, leading to an edge with a dilepton invariant mass $m_{ll} \simeq 78$~GeV.  In both scenarios, one of the pair produced s-bottoms decays to a b-jet and the LSP, and the other decays to a b-jet and a heavier neutralino, which further decays to the LSP and a pair of SFOS lepton through an on-shell s-lepton or an off-shell Z, respectively. Both scenarios feature an edge around 78 GeV and produce a number of events consistent with the observation of CMS. The s-bottom, s-lepton and neutralino masses are within the current experimental limits. 

We showed that the predicted value of the  muon anomalous magnetic moment may be consistent with experiments in both scenarios, what defines additional constraints on the parameters of the model. Moreover, consistency
of the predicted value of the Dark Matter relic density with observations may be obtained and the predicted values of the spin independent Direct Dark Matter cross section are beyond the current limits but may be probed by future searches.    

For simplicity,  we have assumed s-taus to be heavy in both scenarios.  It would be natural to expect the presence of light s-taus in scenario A. In the presence of a light s-tau, we would expect 2b-jets+2  taus + $E_T^{\rm miss}$ as well as a small amount of  2-bjets + 2 taus + SFOS leptons +$E_{T}^{\rm miss}$ and 2 b-jets, 4 taus and $E_T^{\rm miss}$ events at the LHC.  While we don't expect that currently these channels set any additional constraint on this scenario, they must be studied and in the presence of light s-taus they may provide additional ways of testing this scenario at the LHC. A light s-tau would also induce a new Dark Matter annihilation channel and, depending on its mass, could change the discussion we had on this issue, but the discussion on muon anomalous magnetic moment would stay the same (for a recent study on light stau contributions to Dark Matter annihilation, see for instance Ref.~\cite{Pierce:2013rda}).

These scenarios should be probed by a similar analysis of the 8 TeV LHC data from ATLAS. If the existence of the edge is confirmed, further analyses in the 
2~b-jets plus $E_T^{\rm miss}$ channel,  the 4 lepton  plus $E_T^{\rm miss}$ channel, and the $\ge$ 2-bjets $\ge$ 2jets $E_T^{\rm miss}$ channel  in the next run of the LHC will be able to further probe these scenarios.
 
\section{acknowledgments}
We would like to thank N. Craig, C. Hill, B. Hooberman, G. Landsberg, A. Paramonov, L.T. Wang and D. Whiteson for useful discussions. We would also like to thank the Aspen Center for Physics, where this work was started and part of this work has been done. Work is supported by the U.S. Department of Energy under Contract No. DE-FG02-13ER41958. Work at ANL is supported in part by the U.S. Department of Energy under Contract No. DE-AC02-06CH11357.  P.H. is partially supported by U.S. Department of Energy Grant DE-FG02-04ER41286.

\bibliographystyle{utphys}
\bibliography{edgerefs}
\end{document}